\documentclass[conference]{IEEEtran}
\IEEEoverridecommandlockouts
\usepackage{cite}
\usepackage{amsmath,amssymb,amsfonts}
\usepackage{algorithmic}
\usepackage{graphicx}
\usepackage{textcomp}
\usepackage{xcolor}
\usepackage{tabularx, tabularray}
\usepackage{svg}
\usepackage{booktabs}
\usepackage{float}
\usepackage{multirow, enumitem}
\def\BibTeX{{\rm B\kern-.05em{\sc i\kern-.025em b}\kern-.08em
    T\kern-.1667em\lower.7ex\hbox{E}\kern-.125emX}}
\usepackage{float}
\begin{document}

\title{On the Difficulty of Token-Level Modeling of Dysfluency and Fluency Shaping Artifacts}

\author{
\IEEEauthorblockN{Kashaf Gulzar\IEEEauthorrefmark{1}, Dominik Wagner\IEEEauthorrefmark{1}, Sebastian~P. Bayerl\IEEEauthorrefmark{2}, Florian Hönig\IEEEauthorrefmark{3}, Tobias Bocklet\IEEEauthorrefmark{1}, \\ Korbinian Riedhammer\IEEEauthorrefmark{1}}
\\
\IEEEauthorblockA{\IEEEauthorrefmark{1}\textit{Technische Hochschule Nürnberg Georg Simon Ohm}, Germany \\
\IEEEauthorrefmark{2}\textit{Technische Hochschule Rosenheim, Germany}\\
\IEEEauthorrefmark{3}\textit{KST Institut GmbH, Germany}\\
kashaf.gulzar@th-nuernberg.de}
}


\maketitle

\begin{abstract}
Automatic transcription of stuttered speech remains a challenge, even for modern end-to-end (E2E) automatic speech recognition (ASR) frameworks. 
Dysfluencies and fluency-shaping artifacts are often overlooked, resulting in non-verbatim transcriptions with limited clinical and research value. 
We propose a parameter-efficient adaptation method to decode dysfluencies and fluency modifications as special tokens within transcriptions, evaluated on simulated (LibriStutter, English) and natural (KSoF, German) stuttered speech datasets. 
To mitigate ASR performance disparities and bias towards English, we introduce a multi-step fine-tuning strategy with language-adaptive pretraining. 
Tokenization analysis further highlights the tokenizer's English-centric bias, which poses challenges for improving performance on German data. 
Our findings demonstrate the effectiveness of lightweight adaptation techniques for dysfluency-aware ASR while exposing key limitations in multilingual E2E systems.
\end{abstract}


\section{Introduction}
In recent years, end-to-end (E2E) automatic speech recognition (ASR) systems have achieved remarkable performance on fluent speech with up to 95\% word accuracy \cite{tobin2022personalized}. 
However, these large ASR models continue to struggle with stuttered speech with high word error rates (WER) and poor handling of dysfluency events \cite{wu2023world}. 
Unlike typical spontaneous speech, stuttering is characterized by sound and word repetitions, prolongations, silent blocks, and interjections \cite{wu2023world, mujtaba2024inclusive}. 
ASR models are trained on large amounts of fluent and well-structured speech, expecting smooth transitions between phonetic units and grammatically plausible token sequences.
Dysfluency patterns disrupt the statistical and structural assumptions these models rely on, leading to alignment issues, unstable decoding and recognition errors \cite{wu2023world, ma2025asr}.

Moreover, individuals receiving speech therapy frequently adopt fluency shaping techniques designed to suppress dysfluencies.
These include altered articulation patterns such as soft onsets, prolonged phonation and continuous airflow \cite{MALLARD1982fluency}. 
While effective clinically, this modified speech introduces atypical acoustic and prosodic cues that further challenge standard ASR pipelines.

Research addressing stuttered speech in ASR can be broadly categorized into two areas. 
Therapy-related work focuses on detecting, classifying, and analyzing dysfluency events to support clinical assessment and therapy outcomes \cite{amann2024augmenting}.
In contrast, accessibility-oriented work aims to improve WER and downstream performance in tasks such as voice assistants and transcription systems, typically by adapting ASR models or post-processing outputs \cite{mujtaba2024inclusive}.

Early systems extended hidden Markov model (HMM) topologies to model repetitions, prolongations and silent blocks during decoding \cite{noth00_icslp}.
Subsequently weighted finite state transducer (WFST) based DNN-HMM systems improved dysfluent speech handling by incorporating sub-word modeling into the decoding graph, enabling better recognition of partial or fragmented words \cite{smit17_interspeech}.
A widely used modern approach detects dysfluencies by post-processing ASR outputs as a sequence labeling task \cite{chen2022teaching, rocholl2021disfluency, rohanian2021best, omachi2022non}
Alternatively, recent E2E ASR architectures such as RNN-T and Whisper have explored joint modeling strategies, predicting both transcriptions and dysfluencies within a unified framework \cite{kourkounakis2021fluentnet, futami2023streaming, horii2021end, lian2023unconstrained, venkatasubramaniam2023end, mujtaba2024inclusive}.
Additionally, other works have fine-tuned pre-trained ASR models to better recognize unfinished or partial words, further enhancing recognition accuracy on dysfluent speech \cite{ma2023adapting}.
Wagner et al. leverage parameter-efficient fine-tuning through low-rank adaptation (LoRA), which preserves more of the model’s original knowledge \cite{biderman2024loraless} and has shown strong performance in adapting large language models to classifying stuttering in speech segments~\cite{wagner2024large}.
While these approaches typically detect dysfluencies at the transcript level, they often neglect precise temporal information (location), which is critical for applications like speech therapy or conversational analysis \cite{norman2021studying}.

Despite these advances, progress in dysfluency modeling remains limited due to lack of a well-defined problem formulation and high-quality, token-level annotated dysfluency and fluency shaping modifications data especially in multilingual contexts. 
In this work, we address these limitations by integrating transcriptions and dysfluency detection within a token-level modeling framework that can capture standard speech tokens and markers for dysfluencies and fluency shaping artifacts.
To this end, we conduct our experiments on two publicly available datasets: LibriStutter (LSS) and the Kassel State of Fluency (KSoF) datasets, which contain annotated instances of dysfluencies and fluency modifications in English and German. 
Our contributions are:
\begin{itemize}
    \item We perform parameter-efficient fine-tuning of ASR for token-level modeling of dysfluencies and fluency shaping artifacts on LSS and KSoF.
    \item We show that multi-step fine-tuning can reduce language-specific ASR biases and improve recognition performance across languages.
    \item We explore multi-task learning for ASR by modeling fluency-shaping as token-level binary classification, capturing acoustic modifications that can not be represented by discrete tokens.
    \item We identify English-centric tokenization bias that limits recognition accuracy in multilingual ASR.
\end{itemize}

\section{Data}
The LibriStutter (LSS) dataset is a synthetic derivative of 20 hours of the \texttt{dev-clean-100} partition of LibriSpeech (LS), featuring simulated stuttering events \cite{kourkounakis2021fluentnet}.
However, LSS transcripts are non-verbatim, inserting a \texttt{STUTTER} token (along with its type and duration) where a stuttering event occurs.
For consistency, transcriptions were pre-processed by replacing each \texttt{STUTTER} token with a \texttt{<d>} marker at the corresponding word position (cf.~Table~\ref{tab:data}).

The Kassel State of Fluency (KSoF) dataset contains around 5,500 three-second segments recorded during speech therapy sessions in Germany, capturing both natural dysfluencies and fluency shaped speech \cite{bayerl2022ksof}.
In this work, we use the utterance-level segmented version of the complete dataset.
Transcriptions are meticulously annotated by a speech therapist, providing verbatim text with precise timestamps for dysfluencies and modifications.
The dataset comprised of 1,446 utterances, of which 586 (40.5\%) contained at least one dysfluency token and 509 (35.2\%) contained at least one modified speech token.
In pre-processing, the transcriptions were forced-aligned to audio using a DNN-HMM system and time-based markers were inserted into the text as special tokens placed after the corresponding word: \texttt{<d>} indicating a dysfluency event and \texttt{<m>} marking fluency-shaped speech (cf.~Table~\ref{tab:data}).

The German partition of Voxpopuli (VP) dataset contains 200 hours of formal, naturally fluent European parliamentary speeches \cite{wang-etal-2021-voxpopuli}.
Despite being composed of fluent speech, it serves as a valuable resource for fine-tuning ASR models and addressing language-based biases.

\begin{table}[ht]
\centering
\caption{\label{tab:data}
Summary of datasets with example transcriptions after pre-processing.
The \texttt{<d>} and \texttt{<m>} encode the dysfluency and modified speech markers.}
\begin{tabular}{ccl}
\toprule
\multicolumn{1}{c|}{\textbf{Dataset}} & \multicolumn{1}{c||}{\textbf{Type}} & \textbf{Example Transcription} \\
\midrule
\multicolumn{1}{c|}{LSS} & \multicolumn{1}{c||}{\parbox{1cm}{Synthetic}} & \parbox{5cm}{and it may be true replied \texttt{<d>} edward mournfully well} \\
\multicolumn{1}{c|}{KSoF} & \multicolumn{1}{c||}{\parbox{1cm}{Natural}} & \parbox{5cm}{also \texttt{<d>} ja \texttt{<d>} der m- moment \texttt{<m>} der \texttt{<m>} mir} \\
\multicolumn{1}{c|}{VP} & \multicolumn{1}{c||}{\parbox{1cm}{Natural}} & \parbox{5cm}{das beweisen die ergebnisse unserer namentlichen abstimmung} \\
\bottomrule
\end{tabular}
\end{table}

\section{Method}
\begin{figure*}[ht]
    \centering
    \includegraphics[width=\linewidth,
    trim=140 140 140 140,  
    clip
    ]{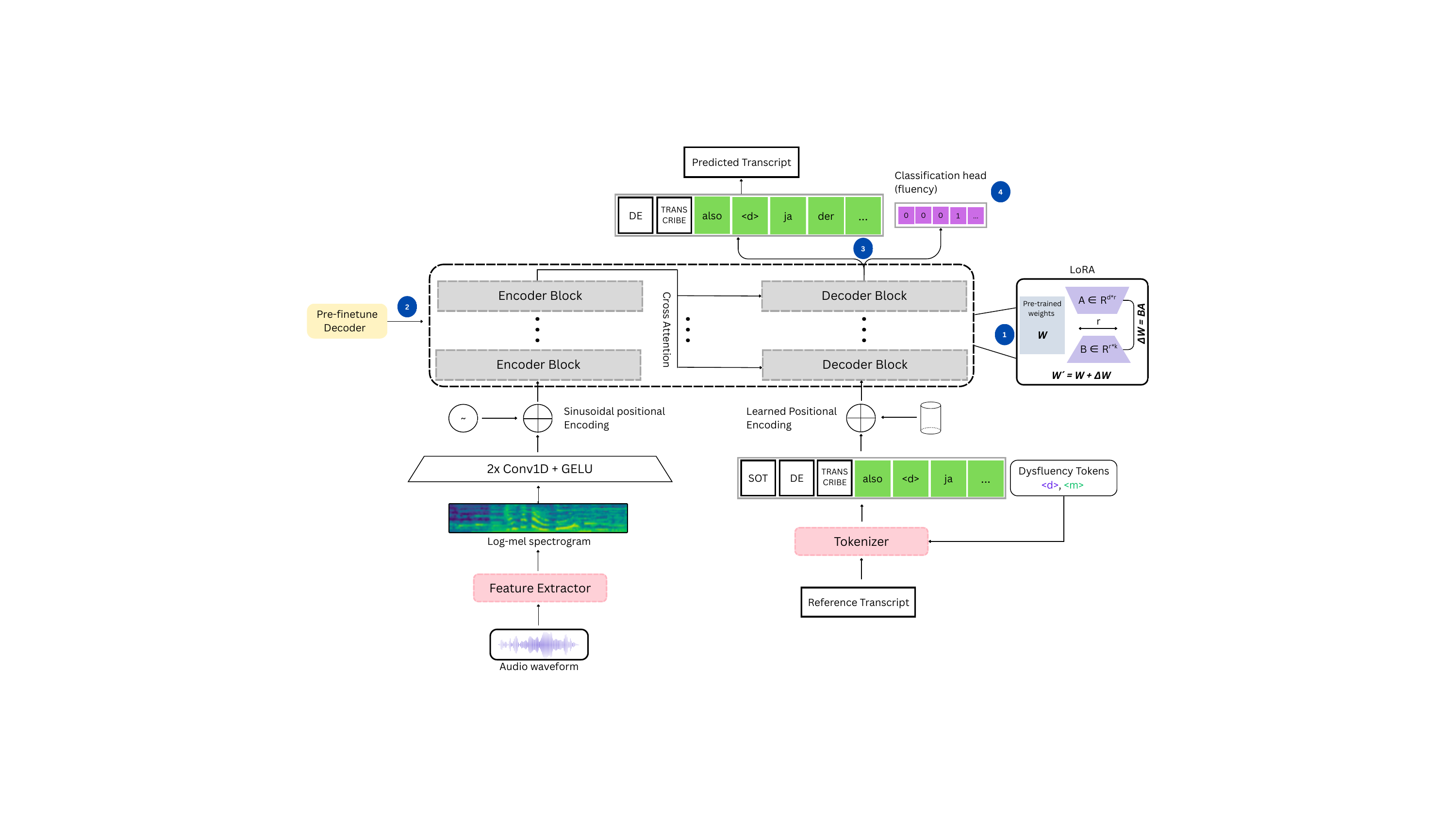}
    \caption{System architecture for parameter-efficient fine-tuning of Whisper using LoRA adapters. Numbered markers indicate experimental variants: (1) LoRA fine-tuning for predicting \texttt{<d>} and \texttt{<m>} tokens; (2) the same, applied to a VoxPopuli-adapted model; (3) simultaneous fine-tuning for \texttt{<d>} prediction and \texttt{<m>} classification with a combined loss; (4) sequential fine-tuning for \texttt{<d>} prediction followed by \texttt{<m>} classification, each with separate losses.}
    \label{fig:architecture}
\end{figure*}

The overall system architecture employs Whisper as the ASR backbone and implements parameter efficient fine-tuning using LoRA as illustrated in Fig.~\ref{fig:architecture}.
Specifically, we utilize \texttt{openai/whisper-large-v3-turbo}, a pruned and fine-tuned 809M-parameter variant of the 1.55B-parameter Whisper V3 model. 
Whisper is a state-of-the-art ASR and speech translation system trained on over 5 million hours of weakly labeled multilingual data \cite{radford2023robust}. 
The \texttt{large-v3-turbo} variant retains the original encoder but reduces the number of decoder layers from 32 to 4, significantly improving inference speed with only a minor degradation in performance.

To achieve further parameter and computational efficiency, we adopt LoRA for fine-tuning. 
LoRA inserts trainable low-rank matrices into the attention and feedforward modules of the pre-trained transformer while keeping the base model parameters frozen \cite{hu2022lora}.
This enables efficient, task-specific adaptation by reducing the number of trainable parameters and the associated computational overhead.
During training, the weights of the pre-trained Whisper model remain frozen and LoRA modules are optimized instead.

Whisper's tokenizer employs byte-level Byte-Pair Encoding (BPE) that converts text into a sequence of subword tokens. 
This strategy offers robustness for multilingual ASR by converting each transcript into a sequence of BPE subword units, where frequently occurring character sequences are represented as individual tokens while rare or out-of-vocabulary words are decomposed into smaller subword fragments \cite{radford2023robust}.
To explicitly model dysfluencies and fluency-shaping artifacts within transcriptions, we extend Whisper's tokenizer by adding two special tokens: \texttt{<d>} for dysfluencies and \texttt{<m>} for fluency-shaped or modified speech. 
These tokens are inserted into the reference transcriptions during pre-processing (cf. Table~\ref{tab:data}) and learned during training of the model.

In all experiments, the model is trained in an auto-regressive manner to predict the next token in the sequence, given the acoustic context, using a cross-entropy loss: 

\begin{equation}
    \mathcal{L} = - \sum_{t=1}^{T} \log P(y_t \mid x_{1:T}, y_{1:t-1}; \theta),
\end{equation}

where $x_{1:T}$ denotes the input acoustic features, $y_{1:T} $ represents the target token sequence (including \texttt{<d>} and \texttt{<m>} tokens), and $\theta$ are the trainable parameters of the encoder.

Fine-tuning is performed for three epochs on a single NVIDIA A100 80GB GPU using 16-bit precision. 
We utilize the Adafactor \cite{shazeer18adafactor} optimizer along with a cosine learning rate schedule \cite{loshchilov2017sgdr} and a peak learning rate of $2 \cdot 10^{-5}$. 
The learning rate is warmed up for 500 steps with evaluation performed at the end of each epoch. 
Greedy decoding is used in all experiments. 
LoRA modules are configured with a rank $r = 64$, a scaling factor for adjusting the magnitude of the adaption $\alpha = 16$, and a dropout probability of 10\%.

\section{Experiments}
The goal of our experiments is to improve transcription accuracy while also enabling a more detailed analysis of speech patterns (dysfluencies and fluency shaping artifacts) relevant to stuttering therapy.
Four distinct experimental configurations are explored (cf~Fig.~\ref{fig:architecture}):
\begin{enumerate}[label=\textbf{(\arabic*)}, leftmargin=0.5cm, itemsep=0.5em]
    \item \textbf{LoRA fine-tuning for \texttt{<d>} and \texttt{<m>} Token Prediction:}
    Tokenizer is extended by adding \texttt{<d>} and \texttt{<m>} tokens. LoRA adapters are then fine-tuned for predicting these special tokens within the transcript.
    \item \textbf{Multi-step fine-tuning:} 
    To address language-based ASR biases, the decoder is first trained on VP German dataset and then LoRA modules are fine-tuned to predict \texttt{<d>} and \texttt{<m>} tokens as in (1). 
\end{enumerate}

At this stage, both dysfluencies and fluency-shaped speech are modeled through token prediction.
However, fluency shaping typically manifests through subtle acoustic modifications rather than distinct lexical units, and thus cannot be fully captured by token-based modeling alone.
Consequently, we differentiate our modeling strategy in experiments 3 and 4: dysfluencies are handled as explicit token predictions within the transcription sequence, but modified speech is modeled via token-level binary classification over the decoder’s hidden representations.

\begin{enumerate}[label=\textbf{(\arabic*)}, leftmargin=0.5cm, itemsep=0.5em]
    \setcounter{enumi}{2}
    \item \textbf{Simultaneous fine-tuning for \texttt{<d>} prediction and \texttt{<m>} classification:} 
    LoRA modules are fine-tuned for \texttt{<d>} token prediction, while auxiliary classification head is added for binary \texttt{<m>} prediction. Both tasks are optimized simultaneously through a combined loss.   
    \item \textbf{Sequential fine-tuning for \texttt{<d>} prediction and \texttt{<m>} classification:} 
    A two-step fine-tuning scheme, where LoRA adapters are first fine-tuned for \texttt{<d>} token prediction within transcripts, followed by a second stage where a classification head is trained with a separate binary loss to predict \texttt{<m>} tokens.
\end{enumerate} 

We train separate models for LSS and KSoF, varying the inclusion of \texttt{<d>} and \texttt{<m>} tokens to analyze their impact on transcription performance. 
Due to differences in speaking style and degree of naturalness between LSS (read speech, synthetic) and KSoF (conversational therapy speech, natural), cross-corpus evaluation was not conducted.
To the best of our knowledge, this is the first study to perform parameter efficient token-level modeling of dysfluencies and binary classification for fluency shaping artifacts in ASR.

\subsection{Token-level Dysfluency and Modified Speech Modeling}
The performance comparison of LoRa fine-tuning for \texttt{<d>} and \texttt{<m>} tokens with and without training on VoxPopuli German dataset for language adaptation is summarized in Table~\ref{tab:results_1_2}.

\begin{table}[h]
\centering
\caption{Token-level dysfluency and modified speech modeling results. Fine-tuning includes tokens representing dysfluencies (\textbf{d}) and/or modified speech (\textbf{m}). $\text{WER}_\text{tok}$ includes dysfluency and modification tokens; results computed using 5-fold cross-validation.}
\label{tab:results_1_2}
\begin{tabular}{ccccc|cc}
\toprule
\multicolumn{1}{c|}{\textbf{Exp.}} & \multicolumn{1}{c|}{\textbf{Train}} & \multicolumn{1}{c|}{\textbf{d}} & \multicolumn{1}{c|}{\textbf{m}} & \multicolumn{1}{c||}{\textbf{Test}} & \multicolumn{1}{c|}{\textbf{WER}} & \textbf{$\text{WER}_\text{tok}$} \\ \midrule
\multicolumn{4}{c|}{\multirow{2}{*}{Baseline}} & \multicolumn{1}{c||}{LSS} & \multicolumn{1}{c|}{0.289} &  \\
\multicolumn{4}{c|}{} & \multicolumn{1}{c||}{KSoF} & \multicolumn{1}{c|}{0.311} &  \\ \midrule
\multicolumn{1}{c|}{\multirow{3}{*}{1}} & \multicolumn{1}{c|}{LSS} & \multicolumn{1}{c|}\checkmark & \multicolumn{1}{c|}{} & \multicolumn{1}{c||}{LSS} & \multicolumn{1}{c|}{\textbf{0.240}} & \textbf{0.232} \\
\multicolumn{1}{c|}{} & \multicolumn{1}{c|}{KSoF} & \multicolumn{1}{c|}\checkmark & \multicolumn{1}{c|}{} & \multicolumn{1}{c||}{KSoF} & \multicolumn{1}{c|}{0.242} & 0.232 \\
\multicolumn{1}{c|}{} & \multicolumn{1}{c|}{KSoF} & \multicolumn{1}{c|}\checkmark & \multicolumn{1}{c|}\checkmark & \multicolumn{1}{c||}{KSoF} & \multicolumn{1}{c|}{0.335} & 0.286 \\ \midrule
\multicolumn{1}{c|}{2a} & \multicolumn{1}{c|}{KSoF} & \multicolumn{1}{c|}\checkmark & \multicolumn{1}{c|}\checkmark & \multicolumn{1}{c||}{KSoF} & \multicolumn{1}{c|}{0.320} & 0.273 \\
\multicolumn{1}{c|}{2b} & \multicolumn{1}{c|}{KSoF} & \multicolumn{1}{c|}\checkmark & \multicolumn{1}{c|}\checkmark & \multicolumn{1}{c||}{KSoF} & \multicolumn{1}{c|}{\textbf{0.285}} & \textbf{0.247} \\ \bottomrule
\end{tabular}
\end{table}

WER and $\text{WER}_\text{tok}$ are reported, where WER is computed on standard transcriptions without dysfluency tokens and $\text{WER}_\text{tok}$ includes the \texttt{<d>} and \texttt{<m>} tokens in both predictions and references. 
All results are averaged over 5-fold cross-validation.

Baseline results demonstrate that the non-finetuned Whisper model struggles on both LSS and KSoF, underscoring the persistent challenges ASR systems face when handling dysfluent and fluency-modified speech.
Fine-tuning with LoRA adapters for dysfluency token prediction (Experiment 1) resulted in substantial WER improvements on both datasets. 
When trained and evaluated on LSS, WER dropped from 0.289 to 0.240, with a corresponding $\text{WER}_\text{tok}$ of 0.232.
A similar pattern was observed for KSoF, where WER reduced from 0.311 to 0.242. 
However, adding both \texttt{<d>} and \texttt{<m>} tokens for KSoF increased WER to 0.335, suggesting that simultaneous modeling of dysfluency and modification tokens introduces additional complexity, affecting transcription accuracy.

Interestingly, multi-step fine-tuning incorporating language adaptation on VoxPopuli German (Experiments 2a and 2b) mitigated this performance drop.  
In Experiment 2a, the Whisper decoder was first fine-tuned on the VoxPopuli German dataset for language adaptation, followed by LoRA fine-tuning on KSoF. 
This setup improved WER to 0.320 and $\text{WER}_\text{tok}$ to 0.273. 
In Experiment 2b, a further enhancement was introduced by fine-tuning the Whisper decoder on VoxPopuli and a subset of KSoF before LoRA fine-tuning on the remaining KSoF set. 
This resulted in the strongest performance, reducing WER to 0.285 and $\text{WER}_\text{tok}$ to 0.247, outperforming both direct fine-tuning (Exp. 1) and the two-stage adaptation in 2a. 

The consistent differences between $\text{WER}_\text{tok}$ and WER suggest that the model recognizes and integrates dysfluency and modification tokens appropriately within transcriptions.
These results confirm the effectiveness of parameter-efficient LoRA fine-tuning for token-level dysfluency modeling and demonstrate that multi-step fine-tuning leveraging language-adaptive pretraining, particularly when incorporating in-domain data early (as in 2b), can improve both transcription and token placement accuracy.

\subsection{Multi-task Learning for Token Prediction and Fluency Classification}
The results of multi-task modeling approaches combining token-level dysfluency prediction with binary classification of modified speech are presented in Table~\ref{tab:results_3_4}.
$\text{WER}_\text{tok}$ is computed for speech and \texttt{<d>} tokens while, accuracy and F1 scores are reported for modified speech classification. 

\begin{table}[h]
\centering
\caption{Performance comparison of multi-task learning for token prediction and modified speech classification. $\text{WER}_\text{tok}$ = WER including dysfluency tokens, Mod = Modified speech, Acc =  accuracy, F1 = F1 score, Simul = Simultaneous, Seq = Sequential. Results are computed using 5-fold cross validation.}
\label{tab:results_3_4}
\begin{tabular}{cc|c|c|c|c}
\toprule
\multicolumn{1}{c|}{\textbf{Exp.}} & \multicolumn{1}{c||}{\textbf{Config}} & \textbf{WER} & \textbf{$\text{WER}_\text{tok}$} & \textbf{Mod. Acc} & \textbf{Mod. F1} \\ \midrule
\multicolumn{2}{c||}{Baseline} & 0.311 &  & 0.699 & 0.436 \\ \midrule
\multicolumn{1}{c|}{3} & \multicolumn{1}{c||}{Simul.} & 0.743 & 0.778 & \textbf{0.818} & 0.534 \\
\multicolumn{1}{c|}{4} & \multicolumn{1}{c||}{Seq.} & \textbf{0.214} &\textbf{0.227} & \textbf{0.779} & 0.498 \\ \bottomrule
\end{tabular}%
\end{table}

As expected, the non-finetuned Whisper model shows limited performance, with a WER of 0.311, modified speech accuracy of 0.699, and a relatively low F1 score of 0.436, reflecting its inability to reliably detect fluency-shaped modifications.
In Experiment 3, where token prediction and modified speech classification were optimized simultaneously, WER increased substantially to 0.743, suggesting interference between the two tasks during combined optimization. 
However, this configuration achieved the highest modification accuracy of 0.818 and an F1 score of 0.534, indicating improved identification of fluency-shaped segments despite overall transcription degradation.

In contrast, Experiment 4, which employed a sequential fine-tuning strategy, first fine-tuning LoRA adapters for dysfluency token prediction followed by training a classification head for fluency modifications, produced the strongest overall performance. 
WER was reduced to 0.214 and $\text{WER}_\text{tok}$ to 0.227. 
While modification accuracy slightly decreased to 0.779 compared to the simultaneous setup, this sequential approach achieved a more balanced trade-off, with a higher overall transcription accuracy and a F1 score of 0.498 for fluency modification detection.

These results indicate that separating the optimization of token prediction and modification classification tasks helps avoid task interference, ultimately improving both transcription quality and modification detection.

\subsection{Tokenization Bias Analysis}
To better understand the persistent performance gap observed in ASR results, particularly on the German KSoF dataset, we conducted a tokenization bias analysis, summarized in Table~\ref{tab:tokenization_bias}. 
This evaluation aimed to quantify how the underlying tokenizer, presumably predominantly trained on English data, handles German speech containing dysfluencies and modified segments.

\begin{table*}[t]
\centering
\caption{Example transcripts with tokenized outputs and tokenization bias metrics across datasets. Tokenization Error Rate (TER) reports the average number of extra tokens per word, while the Token Overestimation Percentage (TOP) reflects the percentage of excess tokens relative to the total token count. The \texttt{<d>} and \texttt{<m>} encode the dysfluency and modified speech markers.}
\label{tab:tokenization_bias}
\begin{tabular}{cllll}
\toprule
\multicolumn{1}{c|}{\textbf{Dataset}} & \multicolumn{1}{c}{} & \multicolumn{1}{c||}{\textbf{Content}} & \multicolumn{1}{c|}{\textbf{TER}} & \multicolumn{1}{c}{\textbf{TOP {[}\%{]}}} \\ \midrule
\multicolumn{1}{c|}{\multirow{2}{*}{LSS}} & \multicolumn{1}{c|}{Text} & \multicolumn{1}{l||}{and it may be true replied \texttt{<d>} edward mournfully well} & \multicolumn{1}{c|}{\multirow{2}{*}{0.106 (0.097)}} & \multirow{2}{*}{8.994 (6.676)} \\
\multicolumn{1}{c|}{} & \multicolumn{1}{c|}{Tokens} & \multicolumn{1}{l||}{\texttt{and}, \texttt{it}, \texttt{may}, \texttt{be}, \texttt{true}, \texttt{replied}, \texttt{<d>}, \texttt{ed}, \texttt{ward}, \texttt{mour}, \texttt{n}, \texttt{fully}, \texttt{well}} & \multicolumn{1}{c|}{} &  \\ \midrule
\multicolumn{1}{c|}{\multirow{2}{*}{KSoF}} & \multicolumn{1}{c|}{Text} & \multicolumn{1}{l||}{ihn \texttt{<d>} ähm \texttt{<d>} so \texttt{<m>} zu sprechen} & \multicolumn{1}{c|}{\multirow{2}{*}{0.389 (0.392)}} & \multirow{2}{*}{23.792 (16.187)} \\
\multicolumn{1}{c|}{} & \multicolumn{1}{c|}{Tokens} & \multicolumn{1}{l||}{\texttt{ihn}, \texttt{<d>}, \texttt{ä}, \texttt{hm}, \texttt{<d>}, \texttt{so}, \texttt{<m>}, \texttt{zu}, \texttt{sprechen}} & \multicolumn{1}{c|}{} &  \\ \bottomrule
\multicolumn{5}{l}{Values are expressed as mean (standard deviation)}\\
\end{tabular}
\end{table*}

To quantify tokenization bias in ASR models, we define two evaluation metrics:

\textbf{Tokenization Error Rate (TER)} measures the number of excess tokens generated relative to the number of words in the reference transcript, indicating the average number of extra tokens produced per word:
\begin{equation}
\text{TER} = \frac{N_{\text{extra tokens}}}{N_{\text{words}}}
\end{equation}

\textbf{Token Overestimation Percentage (TOP)} captures the proportion of all produced tokens that are considered redundant, i.e., additional tokens beyond the expected number of tokens based on the reference:
\begin{equation}
\text{TOP} = \frac{N_{\text{extra tokens}}}{N_{\text{tokens}}} \times 100
\end{equation}
\noindent
where:
\begin{itemize}
    \item $N_{\text{extra tokens}}$ is the number of additional subword tokens generated by the ASR system compared to a verbatim reference.
    \item $N_{\text{words}}$ is the number of words in the reference transcription.
    \item $N_{\text{tokens}}$ is the total number of tokens produced by the ASR model for a given utterance.
\end{itemize}

As expected, the TER and TOP were substantially higher for KSoF compared to LSS dataset. 
Specifically, KSoF exhibited a mean TER of 0.389 and TOP of 23.79\%, compared to 0.106 TER and 8.99\% TOP for LSS. 
This indicates that, on average, the ASR system produces significantly more excess tokens per word for German speech than English speech.
Example transcripts in Table~\ref{tab:tokenization_bias} further illustrate this issue.

While most English words in LSS are preserved as whole tokens, German words, especially those containing dysfluencies (\texttt{<d>}) and characters like umlauts (ß, ä, ö, and ü), or diphtongs (ei, au, eu) are frequently over-segmented into multiple subword tokens-- typically \emph{not} coninciding with phonetic boundaries. 
This behavior reflects the tokenizer's English-centric vocabulary and segmentation patterns, leading to over-fragmentation of German utterances.

These findings explain why fine-tuning alone does not yield substantial WER improvements on KSoF: even with additional task supervision, the underlying tokenization process introduces a structural mismatch that hinders learning and thus accurate transcription. 
This highlights a fundamental limitation in multilingual E2E ASR systems where English-centric tokenizers create persistent barriers for other languages, particularly in specialized domains like fluency disorder therapy.

\section{Discussion}
Our results show that parameter-efficient fine-tuning using LoRA adapters, with dysfluency and modified speech tokens enhances ASR performance on both LSS and KSoF.
While Wagner et al.~\cite{wagner2024large} demonstrated the effectiveness of LoRA for adapting large language models to classify disfluent segments, our findings extend this to multilingual E2E ASR, highlighting its potential for token-level dysfluency-aware transcription in an event-based manner.
Improvements in $\text{WER}_\text{tok}$ suggest the model can learn to recognize and correctly position these specialized tokens, though overall WER improvements on KSoF remain limited.

The multi-task learning experiments indicate that sequential optimization is more effective than simultaneous, likely due to the complexity of handling spontaneous speech and multiple objectives within a single decoding pass.

Tokenization analysis reveals a deeper limitation: the Whisper tokenizer’s English-centric design leads to over-segmentation of German utterances, particularly around dysfluencies. This persistent mismatch between tokenization and language structure undermines transcription accuracy, even after fine-tuning.

Addressing these issues will require tokenization-aware modeling strategies and more adaptable, parameter-efficient fine-tuning approaches.
For instance, Liu et al. proposed a parameter-efficient language extension framework for multilingual ASR, which could be explored in this context~\cite{liu2024parameter}.
Focusing updates on dysfluency tokens or integrating language-specific tokenizers could help overcome current barriers in multilingual ASR for specialized clinical domains.

\section{Conclusion}
We explored the challenges of token-level modeling for dysfluency and fluency-shaping artifacts in multilingual ASR systems. By introducing parameter-efficient fine-tuning with dysfluency and modified speech tokens, we improved event detection and token placement accuracy in both synthetic and natural stuttered speech, particularly reflected in $\text{WER}_\text{tok}$ reductions. However, overall WER improvements, especially on spontaneous German data remained limited.

Through tokenization analysis, we identified an English-centric bias in the Whisper tokenizer, contributing to over-segmentation and persistent transcription errors in German speech. 
These findings underscore a critical limitation in current multilingual E2E ASR frameworks, where tokenization design affects downstream performance in language and domain-specific contexts.

Future work should address these challenges through tokenization-aware modeling, language-specific or adaptive tokenizers, and refined parameter-efficient fine-tuning strategies to better support dysfluency-aware ASR in clinical and multilingual settings.

\bibliographystyle{IEEEtran}
\bibliography{literature}
\vspace{12pt}
\end{document}